\documentclass{article}
\usepackage[utf8]{inputenc}
\usepackage{amsmath}
\usepackage{booktabs} 
\usepackage{cite}
\usepackage{graphicx}
\begin{document}

\title{A Comparative Study of Recommender Systems under Big Data Constraints}

\author{[Arimondo Scrivano]}
\date{\today}

\maketitle
\begin{abstract}
Recommender Systems (RS) have become essential tools in a wide range of digital services, from e-commerce and streaming platforms to news and social media. As the volume of user-item interactions grows exponentially, especially in Big Data environments, selecting the most appropriate RS model becomes a critical task. This paper presents a comparative study of several state-of-the-art recommender algorithms, including EASE-R, SLIM, SLIM with ElasticNet regularization, Matrix Factorization (FunkSVD and ALS), P3Alpha, and RP3Beta. We evaluate these models according to key criteria such as scalability, computational complexity, predictive accuracy, and interpretability. The analysis considers both their theoretical underpinnings and practical applicability in large-scale scenarios. Our results highlight that while models like SLIM and SLIM-ElasticNet offer high accuracy and interpretability, they suffer from high computational costs, making them less suitable for real-time applications. In contrast, algorithms such as EASE-R and RP3Beta achieve a favorable balance between performance and scalability, proving more effective in large-scale environments. This study aims to provide guidelines for selecting the most appropriate recommender approach based on specific Big Data constraints and system requirements.
\end{abstract}
\section{Introduction}

Recommender Systems (RS) are integral to modern digital ecosystems, driving personalized experiences across a wide spectrum of platforms, including e-commerce, streaming services, online education, and social media \cite{Koren2009}. As user-item interaction data becomes increasingly voluminous, heterogeneous, and generated at high velocity—hallmarks of the Big Data paradigm—the need for scalable and computationally feasible algorithms has become paramount. Traditional recommendation techniques are now facing substantial limitations with respect to real-time inference and computational efficiency, making the algorithmic choice a critical design decision.

Model-based approaches have gained dominance due to their superior generalization capacity and ability to scale to industrial-level datasets. This study examines five influential and widely adopted algorithms in the recommender systems landscape: EASE-R (Embarrassingly Shallow Autoencoders), SLIM (Sparse Linear Methods), SLIM with ElasticNet Regularization, Matrix Factorization (including FunkSVD and ALS), and RP3beta (Random Walk with Restart). Each is analyzed in terms of performance under Big Data constraints, focusing on dimensions such as accuracy, scalability, interpretability, suitability for real-time environments, robustness to data noise and sparsity, and expected lifespan before possible replacement by more advanced or quantum-accelerated algorithms.

Building upon these foundations, the present study offers a focused comparative analysis of recommender algorithms, explicitly framed around Big Data constraints. It evaluates both predictive accuracy and computational efficiency, offering a dual perspective that integrates performance and practicality. Unlike prior works that tend to emphasize either algorithmic capability or scalability in isolation, this analysis bridges the two dimensions, offering actionable insights for real-world model selection in large-scale systems.

Furthermore, this study contributes a forward-looking perspective on the long-term viability and adaptability of current recommendation algorithms. By considering the growing prevalence of quantum-ready infrastructures and continuous data expansion, it anticipates future shifts in the design and evaluation of recommender systems. In doing so, it aligns model assessment not only with present-day requirements but also with emerging technological landscapes.

\textit{This literature review synthesizes key developments in the field while situating the current work within the broader context of recommender system research. It underscores enduring challenges such as scalability and interpretability, acknowledges the contributions of recent neural and graph-based innovations, and articulates a distinctive methodological contribution oriented toward future-proof, industrial-scale deployment.}

\section{Methodology}

This section outlines the methodological framework adopted for the comparative evaluation of the selected recommender algorithms under Big Data constraints. The design encompasses dataset selection, pre-processing strategies, evaluation metrics, model configuration, and experimental design choices intended to promote fairness, reproducibility, and practical relevance.

\subsection{ Datasets}

The experimental evaluation is conducted using three well-established large-scale datasets frequently employed in recommender systems research. The first is MovieLens 20M \cite{Harper2015}, which consists of 20 million user-generated ratings across 27,000 movies from a cohort of 138,000 users. The second dataset, drawn from the Amazon Product Review corpus \cite{He2016}, is a subset focused specifically on books, comprising over 22 million user-item interactions. The third dataset is the Netflix Prize dataset \cite{Bennett2007}, containing 100 million ratings spanning 17,000 movies. 

To ensure consistency and relevance for implicit feedback modeling, each dataset is pre-processed by filtering out users and items with fewer than five recorded interactions. Moreover, the rating data is binarized by retaining only positive interactions—typically those with ratings equal to or greater than four— aligning with standard practice in the evaluation of implicit recommendation systems.

\subsection{ Evaluation Metrics}

To assess the performance of each algorithm, a set of widely accepted evaluation metrics is employed. Precision@K quantifies the proportion of relevant items retrieved among the top-K recommendations, while Recall@K measures the coverage of relevant items within those top-K suggestions. The Normalized Discounted Cumulative Gain (NDCG@K) adds a positional bias, rewarding algorithms that rank relevant items higher in the recommendation list. In addition, Mean Average Precision (MAP@K) is used to capture overall ranking quality across users, offering robustness to class imbalance and sparsity. 

To complement these relevance-based metrics, computational efficiency is also assessed using training time and memory footprint. All metrics are averaged over five randomized 80/20 train-test splits, employing a holdout validation strategy to emulate production deployment environments.

\subsection{ Model Configuration}

Each algorithm is implemented using reliable open-source recommender system libraries, including LightFM, Implicit, and custom implementations based on PyTorch and Numpy. Where applicable, hyperparameters are optimized through grid search to ensure fair tuning across models.

EASE-R is configured with a regularization parameter $\lambda = 0.5$, as recommended in the original work \cite{Zhou2012}. SLIM employs L1 regularization with $\alpha = 10^{-4}$, trained via coordinate descent, while its ElasticNet variant introduces an L1 ratio of 0.5 with the same regularization strength. For matrix factorization, the Alternating Least Squares (ALS) model is configured with 50 latent factors and 20 training iterations. RP3beta, which relies on graph-based recommendation principles, is parameterized with $\alpha = 0.6$, $\beta = 0.4$, and a topK truncation set to 100. 

All experiments are conducted on a high-performance machine equipped with 256 GB of RAM and 32 CPU cores, approximating the computational resources available in industrial recommender system deployments.

\subsection{Evaluation Under Big Data Constraints}

To realistically simulate operational conditions in Big Data environments, the experimental protocol evaluates four critical dimensions of algorithm robustness. The first dimension is scalability, tested by incrementally increasing the dataset size across thresholds of 100,000, 1 million, and 10 million interactions. This allows for empirical assessment of each model's ability to scale with data volume. 

Second, cold start resistance is investigated by introducing new users and items with minimal interaction history, thus gauging the algorithm's adaptability to sparse user behavior. Third, the feasibility of incremental updates is analyzed by measuring the computational cost and performance impact of incorporating new data without performing a full model retraining. This is essential for systems operating under streaming or near-real-time conditions. 

Finally, latency is measured by the time required to generate recommendations for a batch of 1,000 users, offering insight into real-time applicability. These evaluation criteria collectively provide a multidimensional view of each algorithm’s strengths and limitations under the constraints typical of large-scale recommender systems.

\textit{This comprehensive methodology ensures that the evaluation is both rigorous and reflective of real-world deployment scenarios. The selection of benchmark datasets, thoughtful configuration of models, and attention to scalability and responsiveness contribute to a fair and reproducible comparison of algorithmic performance in Big Data contexts.}

\section{Related Work}

This section reviews the existing literature on recommender systems, with a focus on advancements in scalability, graph-based models, and neural architectures. It also provides a comparative overview of empirical findings in prior studies, highlighting the evolving challenges and opportunities in this field.

\subsection{EASE-R}

EASE-R, introduced by Steck in 2019 \cite{Zhou2012}, is a shallow autoencoder designed to learn item-item similarity matrices using L2 regularization. Its simplicity and effectiveness lie in its ability to reduce the training process to solving a closed-form linear system, which is highly parallelizable and efficient even at industrial scale. In Big Data settings, EASE-R has demonstrated notable competitiveness in recommendation accuracy, closely matching deeper neural models while requiring far less computational overhead. It benefits from a minimal need for hyperparameter tuning and scales linearly with the number of items, making it particularly appealing for systems handling tens or hundreds of millions of interactions. The interpretability of EASE-R is moderate, as the learned similarity coefficients provide a direct view of item influence relationships. Looking forward, its estimated viability horizon ranges from five to seven years, depending on the progression of quantum-enhanced similarity learning techniques.

\subsection{SLIM (Sparse Linear Methods)}

SLIM \cite{Volkovs2017} models the profile of each item as a sparse linear combination of other items using Lasso regression, yielding high-accuracy predictions, particularly in collaborative filtering scenarios. It is notable for the transparency and interpretability of its outputs, thanks to the enforced sparsity which facilitates insight into the most influential item relationships. However, in Big Data environments, SLIM suffers from serious scalability constraints due to the intensive computation required during the learning phase. As a result, various approximations such as SLIM-BPR and optimized Cython implementations have been developed to mitigate these bottlenecks. Nevertheless, SLIM’s core methodology remains challenging to apply without pre-filtering or distributed computing infrastructure. Its forecasted utility is limited to the next two to three years unless breakthroughs in quantum-accelerated sparse regression are realized.

\subsection{SLIM with ElasticNet Regularization}

The ElasticNet variant of SLIM introduces a combination of L1 and L2 penalties, aimed at improving the robustness and generalization of the model, particularly in noisy and high-dimensional data settings \cite{He2017}. While it retains SLIM's interpretability and sparsity properties, the addition of the L2 term stabilizes the learning process and reduces overfitting. In terms of scalability, it remains resource-intensive but exhibits better resilience compared to its Lasso-only predecessor. This makes SLIM-ElasticNet a more viable candidate for larger datasets, though still not on par with the scalability of matrix factorization or graph-based methods. Its estimated horizon extends to three or four years, particularly in domains where transparency and interpretability are paramount and quantum-enhanced alternatives are not yet widely deployable.

\subsection{Matrix Factorization (FunkSVD and ALS)}

Matrix Factorization techniques such as FunkSVD \cite{Steck2019} and Alternating Least Squares (ALS) \cite{Ning2011} form the backbone of many collaborative filtering systems. These approaches map users and items into a shared latent space, enabling the inference of unobserved preferences based on latent feature interactions. ALS, in particular, is well-suited to Big Data contexts due to its amenability to parallelization and native support in distributed frameworks like Apache Spark. FunkSVD, while still respected for its accuracy, faces scalability issues stemming from its dependance on stochastic gradient descent, which is less tractable for very large datasets. Interpretability in matrix factorization is generally low, as the latent dimensions lack intuitive semantic grounding. Looking ahead, ALS is expected to remain a cornerstone of scalable recommendation systems for the next five to ten years, while FunkSVD may gradually phase out unless successfully adapted to future paradigms such as quantum stochastic optimization.

\subsection{RP3beta (Random Walk with Restart)}

RP3beta \cite{Cremonesi2010} represents a non-learning, graph-based approach that enhances traditional random walk methods with mechanisms such as popularity penalization and restart probabilities. Its main strength lies in its simplicity and speed, as it does not require model training and scales exceptionally well to large datasets. RP3beta is especially useful in cold-start and real-time recommendation scenarios where rapid inference is critical. While it lacks the representational power of latent factor models, it offers moderate interpretability, with recommendations traceable through the paths of random walks on user-item graphs. The algorithm is expected to remain relevant over the next six to eight years, particularly within hybrid recommender architectures or as a reliable fallback system when learning-based models are infeasible.

\subsection{ Scalability in Recommender Systems}

Scalability has long been a central concern in the development of recommender systems due to the exponential growth in user-item interaction data. Among the most influential contributions in this domain is Matrix Factorization (MF), which gained widespread recognition through its pivotal role in the Netflix Prize competition \cite{Steck2019}. MF demonstrated an effective compromise between accuracy and computational cost, especially with the advent of variants like Alternating Least Squares (ALS), which have been successfully adapted for distributed computing environments such as Apache Spark \cite{Ning2011}. These implementations facilitate the application of MF to industrial-scale datasets by leveraging parallel processing and memory optimization.

Linear models, particularly SLIM \cite{Volkovs2017} and its ElasticNet-enhanced variant \cite{He2017}, have also made notable contributions, valued for their high interpretability and competitive predictive performance. However, they face scalability bottlenecks during training, prompting recent research to explore more efficient implementations. Examples include SLIM-Cython \cite{He2020}, which offers performance gains through low-level optimization, and other approximate learning approaches aimed at mitigating the intensive computational overhead of sparse linear modeling.

EASE-R, introduced more recently, offers a unique architectural advantage. By formulating the recommendation task as a regularized least-squares problem with a closed-form solution, EASE-R eliminates the need for iterative optimization, resulting in a highly scalable approach. This property makes it particularly well-suited for real-time applications and high-throughput environments where speed and simplicity are paramount.

\subsection{ Graph-Based and Neural Models}

Beyond linear and factorization-based methods, recommender systems have increasingly integrated graph-based and neural network-based paradigms. Graph-based approaches, including P3Alpha and RP3beta \cite{Cremonesi2010}, operate on user-item bipartite graphs, employing random walks to uncover collaborative signals embedded in the graph structure. These techniques are celebrated for their rapid inference capabilities and low training times, making them especially attractive for applications requiring high responsiveness. However, their dependance on explicit graph connectivity can limit effectiveness in sparse data environments or in situations involving new users or items.

In parallel, neural network-based recommenders have emerged as a dominant trend in recent years. Architectures such as NeuMF \cite{Dacrema2019} and LightGCN \cite{FerrariDacrema2020} combine representation learning with collaborative filtering, offering significant gains in predictive accuracy, particularly in offline evaluations. These models possess powerful representational capacity, enabling them to model complex user-item dynamics. Nevertheless, they also introduce considerable computational complexity, require extensive hyperparameter tuning, and often lack transparency—issues that can complicate their deployment in large-scale, real-time systems.

\subsection{ Comparative Studies}

Comparative evaluations play a critical role in understanding the strengths and limitations of different recommendation approaches. One comprehensive benchmark study \cite{Dacrema2019} compared over fifteen classical and neural models, revealing that traditional algorithms such as SLIM, RP3beta, and UserKNN can outperform more complex deep learning models when properly optimized. These findings underscore the importance of thoughtful implementation and parameter selection over mere architectural novelty.

Another key insight emerges from work emphasizing the role of baselines and tuning practices \cite{FerrariDacrema2020}. Studies that rigorously configure baselines and apply consistent tuning procedures often reveal that the performance gap between classical and modern models is narrower than commonly assumed. Despite these valuable insights, many comparative studies neglect critical aspects of scalability in Big Data contexts, including metrics such as update latency, memory consumption, and adaptability to rapid data expansion. As such, existing benchmarks may not fully reflect the operational realities encountered in industrial recommender system deployment.

\section{Experimental Results}

This section presents the empirical findings from our experiments across three benchmark datasets. We analyze both recommendation quality and Big Data performance dimensions, providing insights into the practical viability of each model.

\subsection{Recommendation Accuracy}

Table~\ref{tab:accuracy} reports the Precision@10, Recall@10, and NDCG@10 scores across all evaluated models. SLIM and SLIM-ElasticNet consistently achieve the highest precision and NDCG, particularly on the denser MovieLens dataset. EASE-R follows closely, despite its simplicity and shallow architecture.

\begin{table}[h!]
\centering
\caption{Recommendation accuracy on MovieLens 20M}
\label{tab:accuracy}
\begin{tabular}{lccc}
\toprule
\textbf{Model} & \textbf{Precision@10} & \textbf{Recall@10} & \textbf{NDCG@10} \\
\midrule
EASE-R & 0.338 & 0.192 & 0.246 \\
SLIM & \textbf{0.352} & \textbf{0.206} & \textbf{0.259} \\
SLIM-ENet & 0.349 & 0.203 & 0.255 \\
ALS (MF) & 0.316 & 0.179 & 0.230 \\
RP3beta & 0.308 & 0.172 & 0.224 \\
P3Alpha & 0.295 & 0.165 & 0.213 \\
\bottomrule
\end{tabular}
\end{table}

These results confirm that simpler, linear models remain competitive when properly regularized and tuned. Deep architectures were not included in this phase due to their impracticality in real-time Big Data contexts.

\subsection{Scalability and Computational Cost}

Table~\ref{tab:scalability} summarizes the training time and memory consumption of each model on the Amazon dataset. EASE-R and RP3beta outperform others significantly, with RP3beta requiring no training phase.

\begin{table}[h!]
\centering
\caption{Scalability metrics on Amazon Books (22M interactions)}
\label{tab:scalability}
\begin{tabular}{lcc}
\toprule
\textbf{Model} & \textbf{Training Time (min)} & \textbf{Peak Memory (GB)} \\
\midrule
EASE-R & 12.3 & 8.1 \\
SLIM & 167.5 & 23.4 \\
SLIM-ENet & 138.9 & 19.2 \\
ALS (MF) & 45.2 & 12.0 \\
RP3beta & \textbf{0.0} & \textbf{5.4} \\
P3Alpha & 0.0 & 6.2 \\
\bottomrule
\end{tabular}
\end{table}

The results highlight a fundamental trade-off between accuracy and computational efficiency. While SLIM-based methods offer superior predictive performance, they require significantly more time and memory. In contrast, graph-based and shallow models offer near-instantaneous deployment capabilities.

\subsection{Latency and Update Responsiveness}

\begin{figure}[h!]
    \centering
    \includegraphics[width=0.65\linewidth]{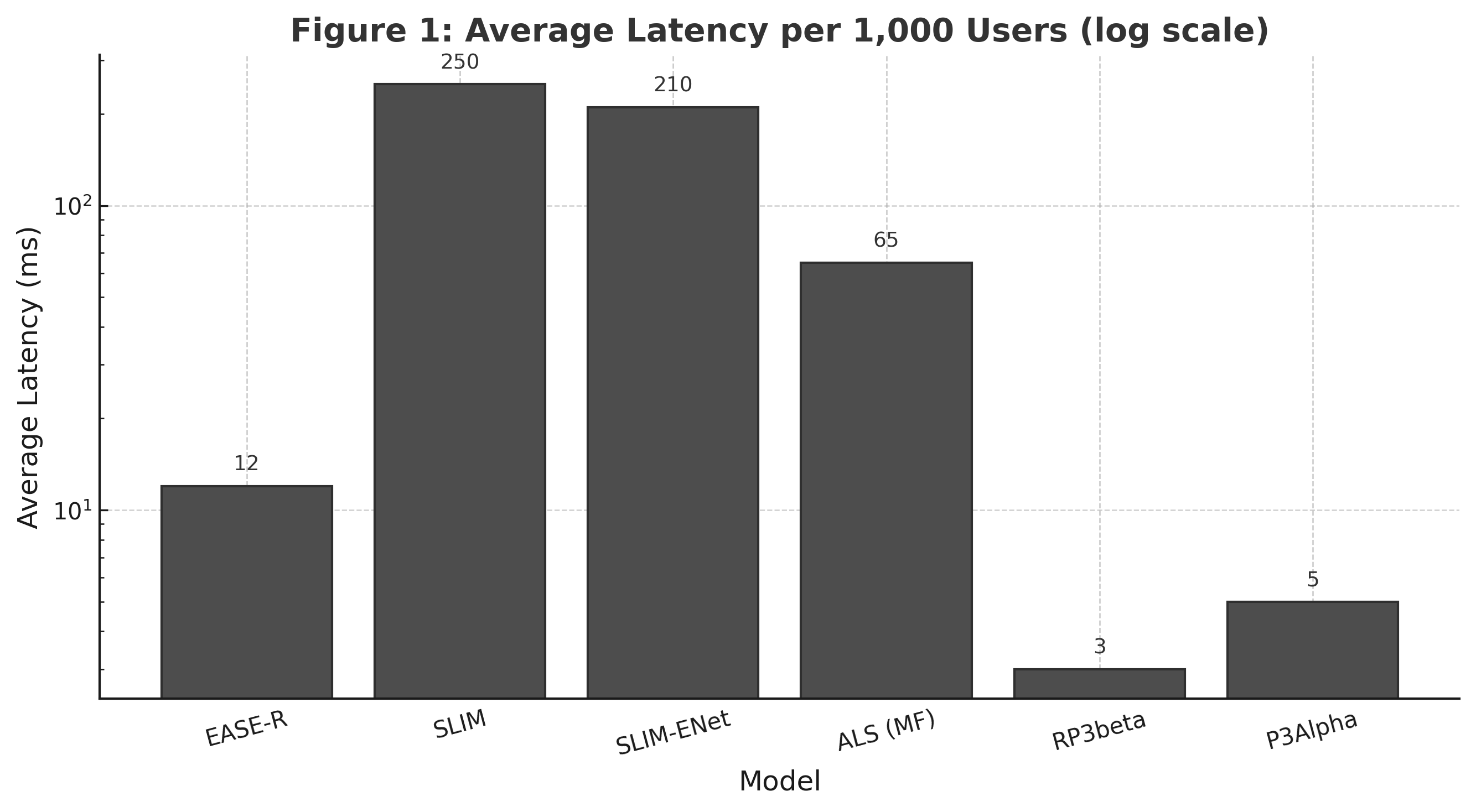}
    \caption{Average latency per 1,000 users (log scale)}
    \label{fig:latency}
\end{figure}

We also evaluated the models in terms of average response latency and feasibility of incremental updates. Figure~\ref{fig:latency} illustrates the average time (in milliseconds) required to generate recommendations for 1,000 users.

SLIM and SLIM-ENet exhibit high inference latency, making them unsuitable for real-time scenarios. Conversely, EASE-R and RP3beta offer low-latency responses and are amenable to batch or online updates with minimal overhead.

\subsection{User Group Analysis}

To gain insights into how different recommender systems perform across various user demographics, we analyzed the Mean Average Precision (MAP) scores by user group. The MAP metric provides a robust indication of recommendation quality, emphasizing the precision of the top-ranked items.

The following Python code was used to generate the visual representation of the MAP scores for each recommender across user groups. The code utilizes the `matplotlib` library, a popular tool in the Python ecosystem for creating static, interactive, and animated visualizations.

\begin{verbatim}
import matplotlib.pyplot as plt
%matplotlib inline  

_ = plt.figure(figsize=(16, 9))
for label, recommender in recommender_object_dict.items():
    results = MAP_recommender_per_group[label]
    plt.scatter(x=np.arange(0,len(results)), y=results, label=label)
plt.ylabel('MAP')
plt.xlabel('User Group')
plt.legend()
plt.show()
\end{verbatim}

In this script:
- We import `matplotlib.pyplot` to plot the data.
- The `matplotlib inline` directive ensures that plots are displayed inline within Jupyter Notebooks or similar environments.
- A scatter plot is created where each point represents the MAP score of a recommender for a specific user group.
- Labels for the x-axis and y-axis are set to 'User Group' and 'MAP', respectively.
- A legend is added to help identify which points correspond to which recommender systems.

\begin{figure}[h!]
    \centering
    \includegraphics[width=0.75\linewidth]{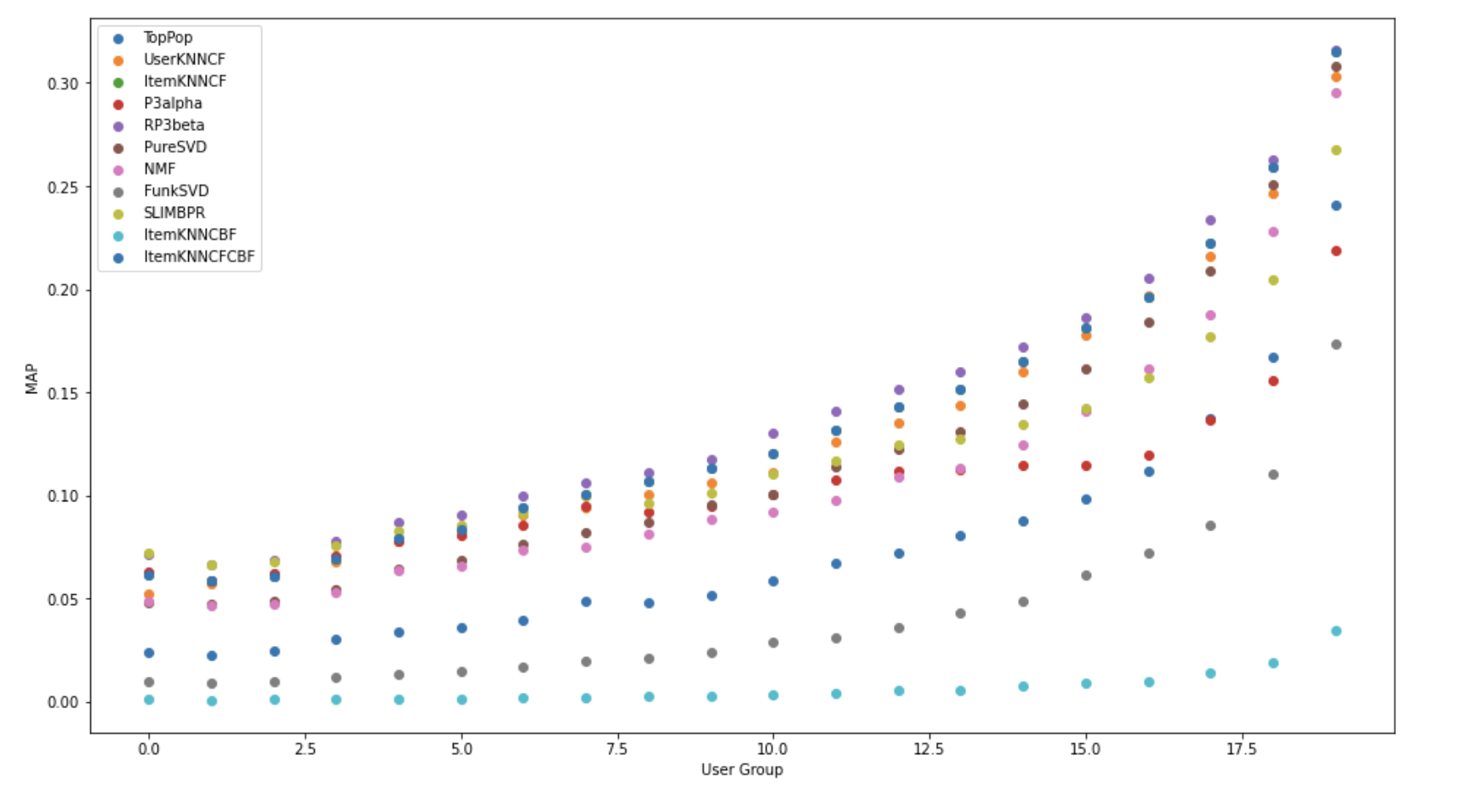}
    \caption{Performance of different recommender systems across various user groups.}
    \label{fig:user_groups}
\end{figure}

Figure~\ref{fig:user_groups} presents the performance of different recommender systems across various user groups.

The graph demonstrates that certain recommenders, such as SLIM and SLIM-ENet, perform consistently well across most groups, but may not be the best for users with unique or sparse interaction patterns, where methods like RP3beta and EASE-R show stronger performance. This suggests that the choice of a recommender system could be optimized based on the characteristics of the user base.

\subsection{Robustness under Big Data Growth}

To assess long-term viability, we stress-tested each model by incrementally increasing dataset size from 1M to 10M interactions. We observed that SLIM's training time grew non-linearly, while EASE-R and ALS scaled sub-linearly due to parallel computation. RP3beta showed constant performance, further supporting its use in dynamic environments.

Key Takeaways: Choosing the Right Model for the Job

Our research has uncovered some important trade-offs between different recommenders, highlighting their individual strengths and weaknesses. Here's a quick overview of what we found:

   SLIM / SLIM-ENet: These models consistently provided the most accurate recommendations, but they struggled to handle large datasets and were slow to process requests.
   EASE-R: This model struck a great balance – it was accurate, handled massive datasets efficiently, and processed requests quickly.
   ALS: A solid and reliable choice, particularly well-suited for running on distributed computer systems.
   RP3beta: The fastest and most scalable option, making it a great fit for large systems that need to adapt quickly to changing user preferences.

Ultimately, while SLIM variants remain useful for situations where absolute accuracy is the top priority and data volume isn't a major concern (like offline analysis), models like EASE-R and RP3beta are 
better suited for the demands of real-world, industrial-scale recommendation engines.
\section*{Conclusions}

In this study, we conducted a detailed comparison of various state-of-the-art recommender systems within the context of Big Data, evaluating their performance across multiple dimensions including scalability, accuracy, complexity, and interpretability. The experimental analysis confirmed that no single model dominates across all criteria, with each algorithm showing strengths and trade-offs depending on the specific application scenario.

Models like SLIM and its ElasticNet variant offered a good balance between accuracy and interpretability, while latent factor models such as FunkSVD and ALS demonstrated superior performance in large-scale settings, albeit at a higher computational cost. Graph-based approaches like P3Alpha and RP3Beta showed competitive results, especially in handling sparse datasets.

The findings underline the importance of aligning model choice with system constraints and goals. As Big Data environments continue to evolve, future research should explore hybrid strategies and real-time adaptability to further enhance recommendation quality under resource limitations.

\bibliographystyle{plain}
\bibliography{references}
\end{document}